\documentclass[aps,twocolumn,showkeys,superscriptaddress]{revtex4-2}
\usepackage{amsmath,amssymb,graphicx}
\usepackage{bm,subfigure,braket,float,pstricks}
\usepackage[colorlinks=true,linkcolor=red,citecolor=blue,urlcolor=blue]{hyperref}
\usepackage{xcolor}

\usepackage{pstricks}
\usepackage{orcidlink}

\begin{document}

\title{Swapped Entanglement in High-Dimensional Quantum Systems}

\author{S. M. Zangi~\!\!\orcidlink{0000-0002-4601-4681}}
\affiliation{Department of Physics, University of Sargodha, Sargodha 40100, Pakistan}

\author{Chitra Shukla~\!\!\orcidlink{0000-0002-6572-8524
}}
\email{ cs@qspacesecuretech.lu}
\affiliation{Interdisciplinary Centre for Security, Reliability and Trust (SnT),\\   University of Luxembourg 1855, Luxembourg}
\affiliation{QSpace SecureTech S.\`a r.l.-S, 19 Rue de l'Industrie, L-8069 Bertrange, Luxembourg}

\author{Khalid Naseer}
\affiliation{Department of Physics, University of Sargodha, Sargodha 40100, Pakistan}

\author{Saeed Haddadi~\!\!\orcidlink{0000-0002-1596-0763}}\email{haddadi@ipm.ir}
\address{School of Particles and Accelerators, Institute for Research in Fundamental Sciences (IPM), P.O. Box 19395-5531, Tehran, Iran}

\begin{abstract}
         Entanglement swapping is a fundamental protocol in quantum information processing that enables the distribution of entanglement between distant quantum systems and plays a central role in quantum communication networks. In this work, we investigate entanglement swapping in arbitrary-dimensional quantum systems (qudits) and provide a quantitative analysis of the resulting entanglement distribution. Using I-concurrence and negativity as entanglement measures, we derive and analyze the average swapped entanglement generated by generalized Bell-state measurements. Our results show that increasing the system dimension enhances the efficiency of entanglement distribution, with high-dimensional systems exhibiting superior performance compared to their qubit counterparts. We further examine the implications of high-dimensional entanglement swapping for long-distance quantum communication and teleportation protocols relevant to quantum repeater architectures. In addition, we study the effects of noise by considering mixed entangled qudit states and analyze the behavior of the swapped entanglement as a function of fidelity and system dimension. The results demonstrate that higher-dimensional systems provide improved robustness against noise, highlighting their potential advantages for future quantum communication technologies.
    \end{abstract}

    \maketitle

    \section{Introduction}
    Quantum entanglement is a distinct feature of quantum mechanics, where two or more particles become interconnected regardless of the distance that separates them \cite{horodecki2009quantum}. This peculiar connection arises when particles interact in specific ways, and their quantum state cannot be described independently for each particle.
    Entanglement forms the foundation of many cutting-edge advancements in quantum science, including quantum teleportation \cite{bennett1993teleporting}, quantum key distribution \cite{ekert1991quantum}, super-dense coding \cite{bennett1992communication}, quantum secret sharing \cite{hillery1999quantum} and quantum secure direct communication \cite{long2007quantum}, etc. This phenomenon also defies classical intuitions about locality and separability, as changes to one particle's state instantly influence its entangled partner \cite{ney2020separability}.
    A particularly fascinating extension of this concept is entanglement swapping \cite{PhysRevLett.71.4287,PhysRevA.57.822}, which demonstrates how entanglement can be transferred between particles that have never interacted directly.

\begin{figure}[b]
        \centering
        \includegraphics[width=0.52\textwidth]{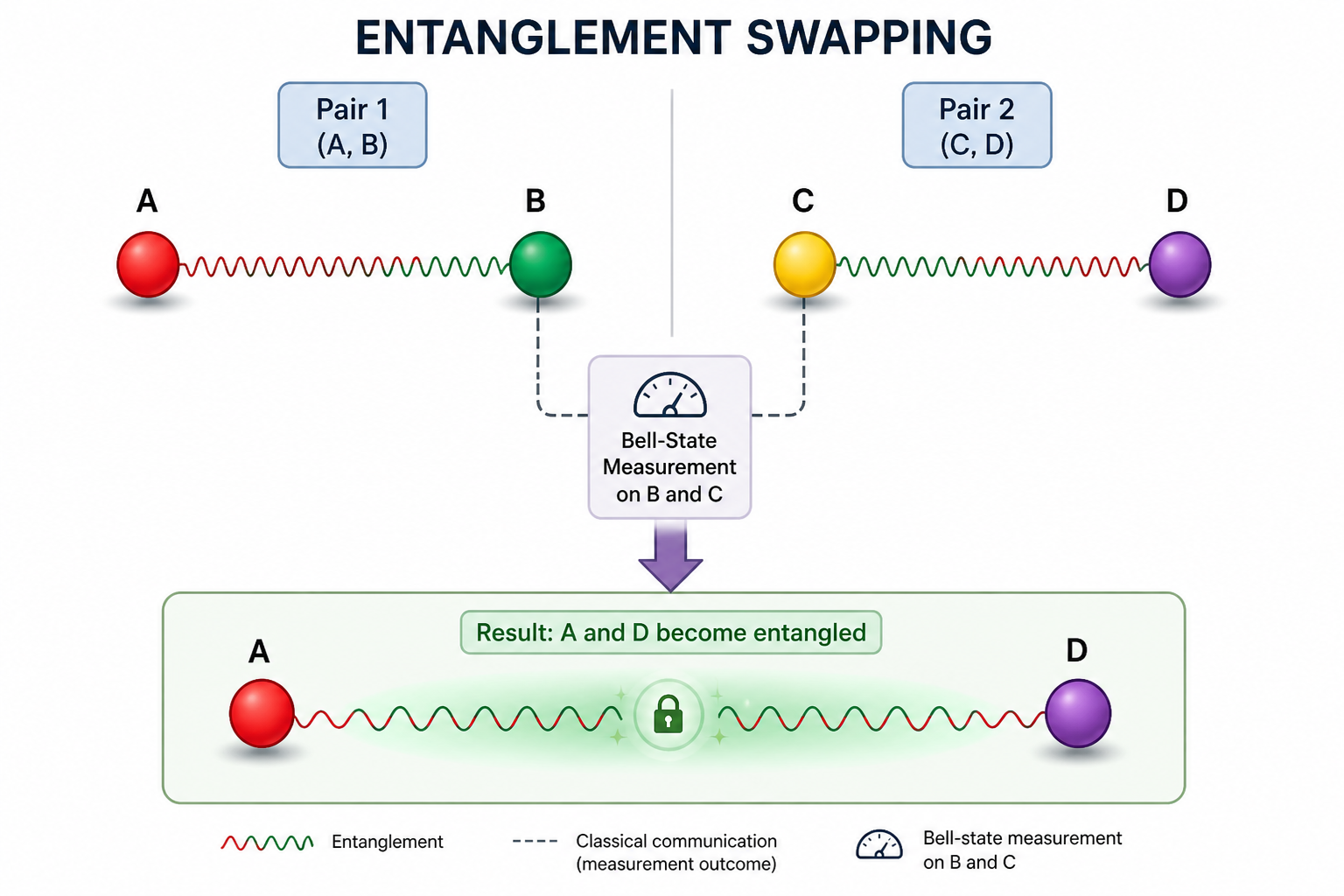}
        \hfill
        \caption{{ The top part of the figure illustrates two independent entangled pairs: (A, B) and (C, D). A Bell-state measurement is performed on particles B and C, leading to entanglement swapping. As a result, unentangled particles A and D become entangled in the bottom part of the figure.}}
        \label{fig:ent-swapping}
        \end{figure}

    Entanglement swapping is an advanced quantum process that creates correlations between particles that were initially unentangled and independent \cite{PhysRevLett.80.3891,zangi2023entanglement}. This is achieved by preparing two pairs of entangled particles, such as A-B and C-D, and then performing a Bell-state measurement on one particle from each pair, such as B and C. This measurement projects the remaining particles, A and D, into an entangled state, even though they have never interacted directly. This process is illustrated in Fig. \ref{fig:ent-swapping}.

    Entanglement swapping is crucial for extending the range of quantum entanglement, making it a cornerstone of quantum communication networks \cite{dur1999quantum}. The linking of entangled segments facilitates the development of quantum repeaters \cite{shchukin2022optimal}. These repeaters are essential for achieving long-distance quantum communication and building scalable quantum internet infrastructure \cite{azuma2023quantum}.

    This work investigates the phenomenon of entanglement swapping in qudit systems~\cite{JanBouda_2001}, emphasizing its implications for quantum networks and high-dimensional quantum communication. Building upon foundational knowledge of entanglement quantifiers such as concurrence and negativity, the study explores the intricate relationships between these measures and fidelity across varying dimensions. It highlights the enhanced efficiency of entanglement swapping in higher dimensions due to the increased degrees of freedom. Furthermore, practical scenarios involving noisy quantum systems are addressed, showcasing methods to maintain and optimize entanglement through techniques like isotropic state analysis. The results advance our understanding of entanglement dynamics in complex quantum systems, offering valuable insights for applications in quantum repeaters and teleportation protocols.

    A growing amount of research indicates that certain techniques are required for the purification and entanglement swapping of quantum systems. Due to the relative sensitivity of the operation and quantum systems, even minor modifications can result in significant changes to the output state.
    Entanglement swapping of starting states into maximally entangled states is discussed by a number of earlier investigations \cite{PhysRevLett.71.4287,PhysRevA.57.822,PhysRevLett.80.3891,zangi2023entanglement}.
    This study on qudit entanglement swapping builds upon and extends prior research on entanglement dynamics and swapping mechanisms in both qubit and qudit systems.
    { General formulations of entanglement swapping for arbitrary initial states and measurements have been developed (see, e.g., the recent paper~\cite{Rosario2025}). In contrast, here we focus on Bell-type measurements and analytically tractable configurations to extract explicit dimension-dependent scaling and operational insights.}

    Although foundational works \cite{zangi2023entanglement,song2014purifying,bergou2021average,karimipour2002entanglement} explored the dynamics of entanglement swapping by focusing on simple relationships and multi-swap scenarios, our paper ventures into the high-dimensional realm of qudits and the quantification of entanglement of final states. Similarly, the study on dissipative systems by Nourmandipour and Tavassoly \cite{nourmandipour2016entanglement} delved into the effects of non-Markovian environments and detuning in qubit-based entanglement swapping. The numerical study of entanglement purification via entanglement swapping among qubit-based systems has been reviewed in Ref. \cite{cong2025formal}. It establishes a formal upper bound on the achievable output concurrence and demonstrates how optimal entanglement purification can be attained.
    In contrast, our work focuses on the practical advantages and challenges of qudits-based systems. We discuss their enhanced capacity for high-dimensional entanglement and resilience to noise, analyzed through measures like I-concurrence \cite{rungta2001universal} and negativity \cite{PhysRevA.65.032314}. By bridging the gap between foundational qubit studies and the complexities of high-dimensional qudits, this paper provides new insights into the potential of qudits for robust and efficient quantum communication and network protocols.

{ The extension to multiple entanglement sources, as relevant for quantum repeater chains, can in principle be implemented via sequential entanglement swapping~\cite{abdrabbou2026}. Although conceptually straightforward, the analytical treatment becomes increasingly complex due to the recursive structure of the protocol. We therefore focus here on the two-source case to obtain closed-form results, leaving the multi-source generalization for future work.
}

    The article is organized as follows. Section \ref{pure-swapping} describes the entanglement swapping among pure qudit systems in detail and the application of the swapped entanglement. Section \ref{noisy-qudits} demonstrates entanglement swapping for noisy qudit systems. The final section \ref{conclusion} contains the concluding remarks.

    \section{Entanglement swapping among qudit systems} \label{pure-swapping}
        With the realization of high-dimensional quantum states \cite{mair2001entanglement,Yurischev2025}, it is of practical importance to study the entanglement swapping among qudit systems.  Suppose there are two $d$-level bipartite pure systems that belong to Hilber spaces $\mathcal{H}_{A}\otimes\mathcal{H}_{B}$   and $\mathcal{H}_{C}\otimes\mathcal{H}_{D}$, respectively.  In the Schmidt form, one can write these states as follows
         \begin{equation}\label{qidits}
        \begin{split}
            \ket{\phi}_{AB} &=\sum_{j_{1}=0}^{d-1}\sqrt{p_{j_{1}}}\ket{j_{1}j_{1}}_{AB}, \\
            \ket{\phi}_{CD} &=\sum_{j_{2}=0}^{d-1}\sqrt{p_{j_{2}}^\prime}\ket{j_{2}j_{2}}_{CD},
        \end{split}
        \end{equation}
        where $\sum_{j_{1}=0}^{d-1}p_{j_{1}}=\sum_{j_{2}=0}^{d-1}p_{j_{2}}^\prime=1$. Let the dimension of $\mathcal{H}_{A}\otimes\mathcal{H}_{B}$ is finite, therefore, we can quantify the entanglement of $\ket{\phi}_{AB}$ by the method developed by Rungta {\it et al.} \cite{rungta2001universal} in terms of I-concurrence as
            \begin{equation}
                C(\ket{\phi}_{AB})=\sqrt{2\left[1-\operatorname{tr}\left(\rho_A^2\right)\right]}=\sqrt{2\left[1-\operatorname{tr}\left(\rho_B^2\right)\right]}.
            \end{equation}
        Also, the negativity \cite{lee2003convex} of the biqudit state $\ket{\phi}_{AB}$ can be obtained as

        \begin{equation}
        \mathcal{N}(\ket{\phi}_{AB}) =\frac{2}{d-1} \sum_{i_{1}<j_{1}} \sqrt{p_{i_{1}} p_{j_{1}}}.\label{neg-Eq3}
        \end{equation}
        Similarly, one can calculate the I-concurrence and negativity for the state $ \ket{\phi}_{CD}$.

        The initial state of our four-qudit system is
        \begin{equation}\label{4-qidits}
            \ket{\Phi}_{ABCD} =\sum_{j_{1},j_{2}=0}^{d-1}\sqrt{p_{j_{1}}p_{j_{2}}^\prime}\ket{j_{1}j_{1}j_{2}j_{2}}_{ABCD}.
        \end{equation}
        Performing a Bell measurement on the particles (B, C), one gets
        \begin{equation}\label{Bell-qidits}
            \ket{\Psi_{uv}}_{BC} =\frac{1}{\sqrt{d}}\sum_{l=0}^{d-1} e^{\frac{2\pi i}{d} l u}\ket{l}_{B}\ket{l\oplus v}_{C},
        \end{equation}
        where $u,v\in \{0,1\dots d-1 \}$ and $\oplus$ denotes modulo $d$ addition. Then, the unentangled particles A and D are projected onto an entangled state
        \begin{equation}\label{AD-qidits}
            \ket{\Phi_{uv}}_{AD} =\frac{1}{\sqrt{P_{uv}}}\sum_{l=0}^{d-1}\sqrt{p_{l} p_{l \oplus v}^{\prime}} e^{\frac{-2\pi i}{d} l u}\ket{l}_{A}\ket{l\oplus v}_{D},
        \end{equation}
        where normalization factor is $ P_{uv} =\sum_{l=0}^{d-1}p_{l} p_{l \oplus v}^{\prime}$. The probability of any outcome state $\ket{\Psi_{uv}}_{BC}$ is $P_{uv}/d$.
         The probabilities of the outcome quantum state and the final entangled state depend on the initial four-qudit state. The probabilities of either state can be calculated by taking the inner product of the joint state $\ket{\Phi}_{ABCD} $ with the desired state.

         The I-concurrence of the state $ \ket{\Phi_{uv}}_{AD}$ is derived by
         \begin{equation}
            C_I\left(\ket{\Phi_{uv}}_{AD}\right)=\left[2\left(1-\frac{1}{(P_{uv})^2} \sum_{l=0}^{d-1}\left(p_l p_{l \oplus v}^{\prime}\right)^2\right)\right]^{1 / 2}.
        \end{equation}
        The average I-concurrence can be obtained by multiplying the above expression with the probability $P_{uv}/d$ and adding the results for all possible outcomes, namely
        \begin{equation}
            C_I^{(\text{av})}=\sqrt{2}  \sum_{v=0}^{d-1}\left[\left(\sum_{l=0}^{d-1}p_{l} p_{l \oplus v}^{\prime}\right)^2-\sum_{l=0}^{d-1}\left(p_l p_{l \oplus v}^{\prime}\right)^2\right]^{1 / 2}.
        \end{equation}
        This measure quantifies entanglement based on the square root of a sum of terms involving the probabilities $p_{l}$ and $p_{l\oplus v}^{\prime}$, where $l$ or $v$ ranges from $0$ to $d-1$.
        The presence of nonzero values of $C_I^{(\text{av})}$ indicates the generation of entanglement between particles A and D.

        The negativity of the state $ \ket{\Phi_{uv}}_{AD}$ is

        \begin{equation}\label{neg-phi}
        \mathcal{N}\left(\ket{\Phi_{uv}}_{AD}\right)=\frac{2}{P_{u v}(d-1)} \sum_{i<j} \sqrt{p_i p_{i \oplus v}^{\prime} p_j p_{j \oplus v}^{\prime}}.
        \end{equation}
         Note that the phase factor $e^{\frac{-2\pi i}{d} l u}$ disappears in I-concurrence and negativity expressions because it cancels out in the density matrix $\left|\Phi_{uv}\right\rangle_{AD}\left\langle\Phi_{uv}\right|$. So, the negativity depends only on $v$ and we can write $\mathcal{N}\left(\left|\Phi_{uv}\right\rangle_{AD}\right)$ as $\mathcal{N}_{v}$.
        Since there are $d^2$ possible Bell measurements on particles A and D, the total number of possible entangled states after swapping is also $d^2$.
        The average negativity of $d^2$ possible final states $\left|\Phi_{uv}\right\rangle_{AD}$ after entanglement swapping can be calculated by

        \begin{equation}\label{neg-ave}
        \begin{aligned}
            \mathcal{N}_{\mathrm{av}}  = \sum_{u, v=0}^{d-1} \frac{P_{u v}}{d} \mathcal{N}_{v}
             =\frac{2}{(d-1)} \sum_{v=0}^{d-1}\sum_{i<j} \sqrt{p_i p_{i \oplus v}^{\prime} p_j p_{j \oplus v}^{\prime}}.
        \end{aligned}
        \end{equation}

        If the value of I-concurrence becomes zero, then entanglement is not present between particles; otherwise, particles are entangled.
        However, this is true for negativity under certain conditions~\cite{haddadiQINP2025}.

        The initial probabilities ($p_j$ and $p_{j}^\prime$) of the Schmidt decomposed states influence the final entanglement between A and D. The specific distribution of these probabilities can lead to varying degrees of entanglement. The success of entanglement swapping also relies on a successful Bell state measurement on particles B and C. The outcome of this measurement determines the specific entangled state shared by A and D.
 \begin{figure}[t]
        \centering
        \includegraphics[width=0.48\textwidth]{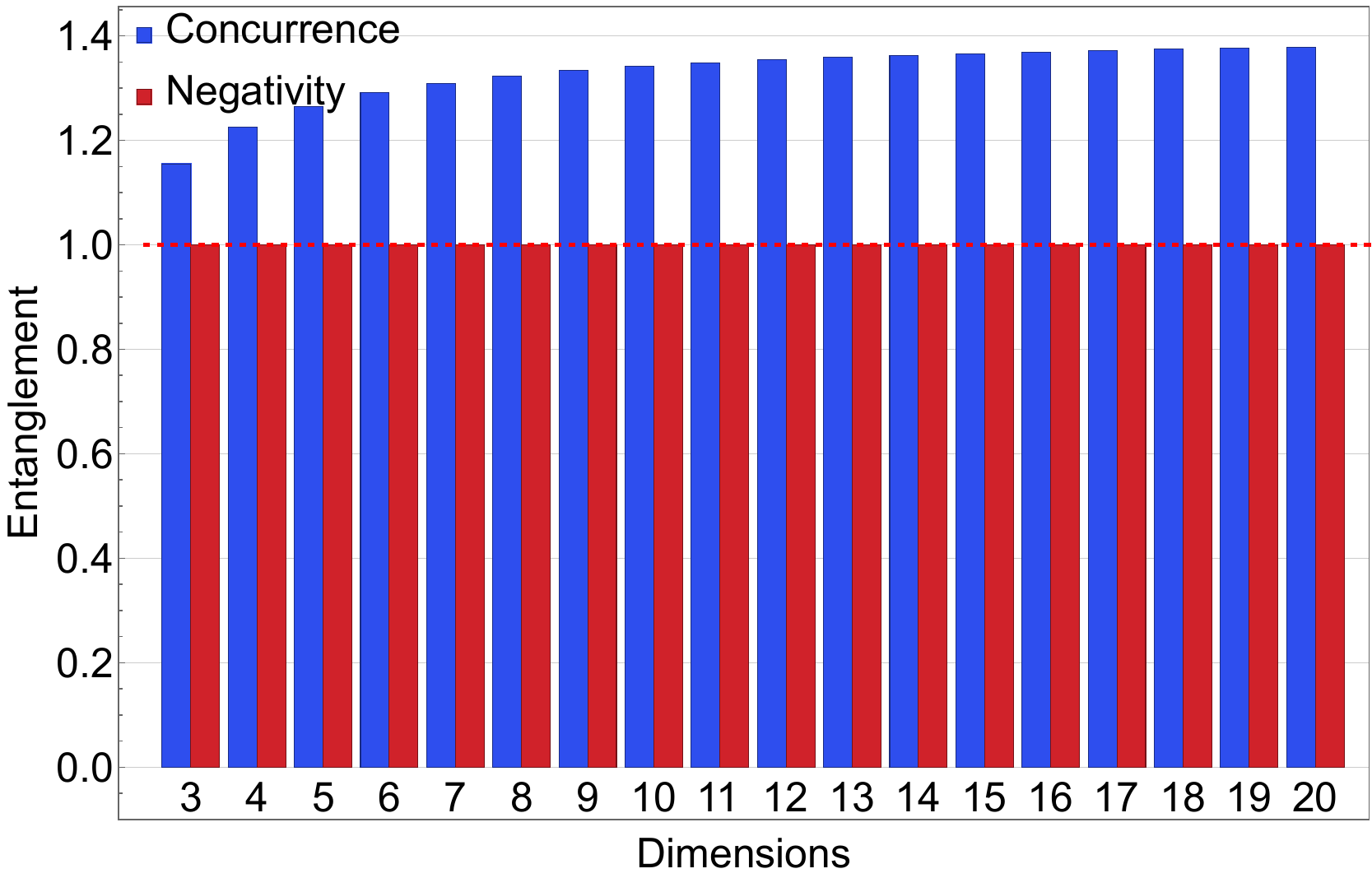}
        \hfill
        \caption{(Color Online) This plot illustrates the average entanglement of the final states of particles AD, which is generated following entanglement swapping, quantified in terms of I-concurrence and negativity.}
        \label{fig:Max-C-N-ave}
        \end{figure}

        Maximally entangled states are characterized by having the highest possible degree of correlation between the constituent parts. If we take both AB and CD states maximally entangled, then we have $p_{j_{1}}=p_{j_{2}}^\prime=1/d$.
        For maximally entangled initial states, entanglement swapping produces a maximally entangled state between particles A and D. Figure~\ref{fig:Max-C-N-ave} shows the average I-concurrence and average negativity of the swapped state as functions of the dimension $d$. The purpose of this figure is not to demonstrate the existence of entanglement in the swapped state, which is already guaranteed by the protocol, but rather to examine how the amount of entanglement generated through the swapping process scales with the Hilbert-space dimension. As shown in the figure, the average I-concurrence, $C_I^{(\text{av})}$, increases with increasing dimension, reflecting the larger entanglement capacity available in higher-dimensional systems. In contrast, the normalized negativity, $\mathcal{N}_{\text{av}}$, remains essentially unchanged with dimension. This difference indicates that the two measures capture distinct aspects of high-dimensional entanglement. Consequently, Fig.~\ref{fig:Max-C-N-ave} provides a useful reference for interpreting the dimensional dependence observed in the subsequent analysis of entanglement swapping and its robustness under noisy conditions.

      {{Note that the negativity used in Eq.~\eqref{neg-Eq3} or Eq.~\eqref{neg-ave} includes a normalization factor $2/(d-1)$ which constrains its maximum value to unity for all dimensions.}} This make it an effective tool for comparing entanglement across different dimensions. However, this normalization also imposes a limitation, as it may not fully capture the increasing complexity and strength of entanglement in higher-dimensional quantum systems.
     On the other hand, I-concurrence is not a normalized measure and exhibits values exceeding 1 as the dimension increases.
     This behavior indicates that {{I-concurrence exhibits explicit dimension dependence and is more sensitive to variations in the Schmidt coefficients in higher-dimensional systems.
     However, this growth should not be directly interpreted as an increase in a physical entanglement resource.}}
     The fact that negativity saturates at one suggests that it may underestimate entanglement in larger dimensions, whereas I-concurrence continues to scale and reflects the richer structure of quantum correlations in qudit-based systems. This makes I-concurrence a more effective entanglement measure because it {{reflects the increasing entanglement capacity of higher-dimensional systems}}. Negativity remains useful for bounded nature, I-concurrence provides a deeper and more accurate understanding of entanglement dynamics in high-dimensional spaces. This makes it a superior choice for analyzing entanglement in  larger quantum systems.

    The final state $|\Phi_{uv}\rangle_{AD}$ has a wide range of applications; e.g. long-distance teleportation can harness the advantage of qudit Bell entangled based quantum repeaters. Specifically, $|\Phi_{uv}\rangle_{AD}$ can be used as quantum repeaters in a non-terrestrial quantum network application to teleport an unknown qudit system from one end to the other end via several intermediate nodes. Before executing qudit teleportation, it is necessary to perform the purification \cite{miguel2018efficient} of $|\Phi_{uv}\rangle_{AD}$ at each node to obtain maximally entangled states for qudit Bell pairs and achieve an efficient qudit teleportation protocol. Considering this scenario, we discuss the qudit teleportation of $|\Phi_{uv}\rangle_{AD}$, obtained in Eq. \eqref{AD-qidits}, after the entanglement swapping. After the purification process of $|\Phi_{uv}\rangle_{AD}$, a maximally entangled qudit Bell pair $|\Phi_{uv}\rangle'_{AD}$ can be obtained, which works as a shared quantum channel between A and D (two nodes).

    Now, Alice, who has node A, wishes to teleport an unknown qudit state $|\phi\rangle=\mathbin{\sum_{j=0}^{d-1}\alpha_{j}}|j\rangle$
to Diana at node D. To proceed, Alice performs a $d^{2}$-outcome joint projective measurement on her qudit A shared with D and the unknown qudit $|\phi\rangle$
to be teleported. She performs a joint projective measurement in a generalized Bell-like basis $\{|\Phi_{uv}\rangle\}$, elements of which can be expressed as
\begin{equation}
|\Phi_{uv}\rangle=\mathbin{\sum_{k=0}^{d-1}\beta_{ku}}|k,k\oplus v\rangle,\label{eq:Qudit_Bell_basis}
\end{equation}
where the symbol \textquotedblleft $\oplus$\textquotedblright{} indicates
sum modulo $d$ and the coefficients $\beta_{ku}=\frac{1}{\sqrt{d}}e^{\frac{2\pi i}{d}k.u}$
 stand for the extent of entanglement which satisfy the
relation $\sum_{k=0}^{d-1}\beta_{ku}\beta_{ku'}^{*}=\delta_{uu'}$.
After Alice's qudit Bell measurement, the combined channel can be written as
\begin{equation}
\left(|\Phi_{uv}\rangle\langle\Phi_{uv}|\otimes\mathbb{I}_{d}\right)|\phi\rangle\langle\phi|\otimes\hat{\rho}_{\text{ch}},\label{eq:Combind_state_after_BM}
\end{equation}
where $\mathbb{I}_{d}$ is the $d$ dimensional identity operator, $|\phi\rangle\langle\phi|$
 shows the density operator of the unknown qudit to be teleported, and $\hat{\rho}_{\text{ch}}$  describes the density operator of the quantum channel shared by Alice and Diana. Alice conveys measurement outcome $(u,v)$ to Diana using an authenticated classical channel. Accordingly, Diana applies a local unitary operation
on her qudit given by one out of the $d^{2}$ Weyl operators $\hat{U}_{uv}$ \cite{fonseca2019high}, defined as
\begin{equation}
\hat{U}_{uv}=\mathbin{\sum_{j=0}^{d-1}\omega_{d}^{ju}}|j\rangle\langle j\oplus v\rangle.\label{Unitary_operator}
\end{equation}
After each iteration, the state of Diana\textquoteright s qudit (up to normalization) can be expressed as
\begin{equation}
\hat{\rho}_{uv}=\hat{U}_{uv}\operatorname{tr}_{A}\left\{ \left(|\Phi_{uv}\rangle\langle\Phi_{uv}|\otimes\mathbb{I}_{d}\right)|\phi\rangle\langle\phi|\otimes\hat{\rho}_{\text{ch}}\right\} \hat{U}^{\dagger}_{uv},\label{eq:Bob_state}
\end{equation}
where $\operatorname{tr}_{A}$ stands for the partial trace on the subsystem of Alice.

While quantum teleportation using quantum entanglement holds great promise, it is highly vulnerable to noise, leading to potential loss of teleported information. However, this challenge can be addressed by leveraging the quantum superposition of different causal orders through a quantum switch \cite{dey2023entanglement}. Specifically, if the entanglement is distributed between the sender and receiver using a quantum switch \cite{dey2023entanglement}, the fidelity of teleportation can be improved even in the presence of noise, enabling a wider range of states to be accurately teleported \cite{mylavarapu2024teleportation}.

    \section{Entanglement Swapping among Noisy Qudits}\label{noisy-qudits}

    Thus far, our investigations have focused on pristine quantum systems, carefully shielded from external influences. However, in practical scenarios, quantum systems inevitably engage with their surroundings, encountering a spectrum of disturbances and fluctuations. Among these, a prominent disturbance is depolarizing noise, also known as white noise. This form of interference transforms the quantum state into a new state by considering a maximally mixed state as $\mathbb{I}_{d}/d$. 
    To illustrate, let us examine a scenario involving a noisy biqudit state $\rho_{AB}^{N}$, { obtained as a convex mixture of a maximally entangled state $\rho_{AB}$ and white noise:}
     \begin{equation}\label{noisyqd}
        \rho_{AB}^{N}=\alpha \rho_{AB}+(1-\alpha)\frac{\mathbb{I}_{d} \otimes \mathbb{I}_{d}}{d^2},
    \end{equation}
    where $\frac{-1}{d^2-1}\leq\alpha\leq 1$ is the visibility of the system AB, and for simplicity, we choose $\rho_{AB}=\ket{\phi}_{AB}\bra{\phi}$ with $\ket{\phi}_{AB}=\sum_{j=0}^{d-1}\frac{1}{\sqrt{d}}\ket{jj}_{AB}$. Equation \eqref{noisyqd} represents the isotropic form \cite{horodecki1999reduction} of the system AB. The isotropic systems remain invariant under all unitary transformations of the form $U\otimes U^{*}$, where the asterisk denotes complex conjugation in a certain basis.

    In Bloch representation, the isotropic states can be reformulated as \cite{yang2021decompositions}:
    \begin{equation}\label{Ibloch}
    \rho_{AB}^{N}=\frac{\mathbb{I}_{d} \otimes \mathbb{I}_{d}}{d^2}+\frac{1}{4} \sum_{\mu=1}^{d^2-1} \frac{2 \alpha}{d} \lambda_\mu \otimes \lambda_\mu^{\mathrm{t}},
    \end{equation}
    where $\lambda_\mu$ are $d^2-1$ generators of SU($d$) group and superscript $t$ represents the transpose operation of a matrix. Now, from Eq. \eqref{Ibloch}, one can construct the Bloch matrix:
    \begin{equation}\label{Bloch_formj}
         \tilde{\mathcal{T}}=
         \left(
            \begin{array}{cc}
                c & \Vec{O}_1 \\
                \Vec{O}_2 & T \\
            \end{array}
        \right),
    \end{equation}
    where $c$ is a scalar number, $T_{\mu \mu}=\operatorname{tr}\left(\rho_{AB}^{N} \lambda_\mu \otimes \lambda_\mu^{\mathrm{t}}\right)$, and $\Vec{O}_1$ and $\Vec{O}_2$ are row and column vectors, respectively, with all entries zero and dimensions $d^2-1$. The combo separability criteria \cite{zangi2021combo} tells us that the isotropic states remain entangled if $ \|\tilde{\mathcal{T}}\|_{\text{KF}}-1>0$, otherwise, become separable.

    Let the particles C and D also be in an isotropic state such that $\rho_{CD}^{N}=\rho_{AB}^{N}$. Now the initial state of a four-qudit noisy system is
    \begin{equation}
        \begin{aligned}
            \varrho^{N} & \equiv \rho_{AB}^{N} \otimes \rho_{CD}^{N} =\left(\alpha\rho_{AB}+\beta \mathcal{I}\right) \otimes\left(\alpha\rho_{CD}+\beta \mathcal{I}\right),
    \end{aligned}
    \end{equation}
    where $\mathcal{I}=\mathbb{I}_{d} \otimes \mathbb{I}_{d}$ and $\beta=(1-\alpha)/{d^2}$. {{ Here, $\alpha$ quantifies the weight of the ideal state, while $\beta$ is fixed by normalization and represents the contribution of white noise.}}

    We can generate entanglement between particles A and D by using the entanglement swapping technique for isotropic states of particles A, B and C, D. For this purpose, we make Bell-state measurement on qudits B and C via a complete set of projective operators $\ket{\Psi_{uv}}_{BC}\bra{\Psi_{uv}}$, developed from Eq. \eqref{Bell-qidits}. After Bell measurement on the qudits B and C, qudits A and D which were initially separable, become entangled in the following form:
    \begin{align}\label{ADnoisyqd}
        \rho_{AD}^{N}=&\frac{\alpha^2}{d}\sum_{i,j=0}^{d-1}\ket{ii}_{AD}\bra{jj}+d^2 \beta^2 \mathcal{I}\nonumber\\
        &+\alpha\beta\left(\sum_{j=0}^{d-1}\ket{j}\bra{j}\otimes \mathbb{I}+ \mathbb{I}\otimes\sum_{j=0}^{d-1}\ket{j}\bra{j}\right)
        .
    \end{align}
    We find $d^2$ output states $\rho_{AD}^{N}$, each with probability $1/d^2$. As both biqudit input states have the same amount of entanglement, all the final states also have an equal amount.

    Using now the local unitary operators, one can transform the final noisy entangled states of the qudits A and D, given in Eq. \eqref{ADnoisyqd}, to the following isotropic form:
    \begin{align}\label{isotropic}
    \rho_{AD}^{(\mu\nu)}=& {\beta^{{\prime}}}\left(\sum_{k=0}^{d-1} e^{-2 \pi i \mu k / d}\left|k\right\rangle\left|k\ominus\nu\right\rangle\right)\nonumber\\
     &\times\left(\sum_{l=0}^{d-1} e^{+2 \pi i \mu l / d}\left\langle{l\ominus\nu} \right|\left\langle l\right|\right)+(1-{\beta^{{\prime}}}) \frac{\mathbb{I}_{d^2}}{d^2},
    \end{align}
    where $\beta^{\prime}=\alpha^{2}$ is the visibility of the entangle`ment between qudits A and D.  Therefore, for all $d^2$ quantum states $\rho_{AD}^{(\mu\nu)}$ the average output I-concurrence $C_I^{(av)}(\rho_{AD}^{(\mu\nu)})$ is the same as that of the I-concurrence of each  $\rho_{AD}^{(\mu\nu)}$ state.

    The I-concurrence \cite{rungta2003concurrence} of each isotropic state $\rho_{AD}^{(\mu\nu)}$ in Eq. \eqref{isotropic} is expressed as
    \begin{equation}\label{iso-conc}
    C_I\left(\rho_{AD}^{(\mu\nu)}\right)= \begin{cases}0, & F \leqslant 1 / d, \\ \sqrt{\frac{2 d }{d-1}}(F-\frac{1}{d}), & 1 / d < F \leqslant 1,\end{cases}
    \end{equation}
    where {$F=\langle\Psi|\rho_{AD}^{(\mu\nu)}| \Psi\rangle=\beta^{\prime}+\frac{1-\beta^{\prime}}{d^2}$}, with $0\leqslant F\leqslant1$, is the \textit{fidelity} of $\rho_{AD}^{(\mu\nu)}$ and $|\Psi\rangle=\sum_{k=0}^{d-1} e^{-2 \pi i \mu k / d}|k\rangle|k\ominus\nu\rangle$. This fidelity is directly proportional to $\beta^{\prime}$, which is the visibility of the swapped entanglement between qudits A and D.
    It means low noise has participated in the swapping process.

\begin{figure}[t]
         \centering
        \includegraphics[width=0.48\textwidth]{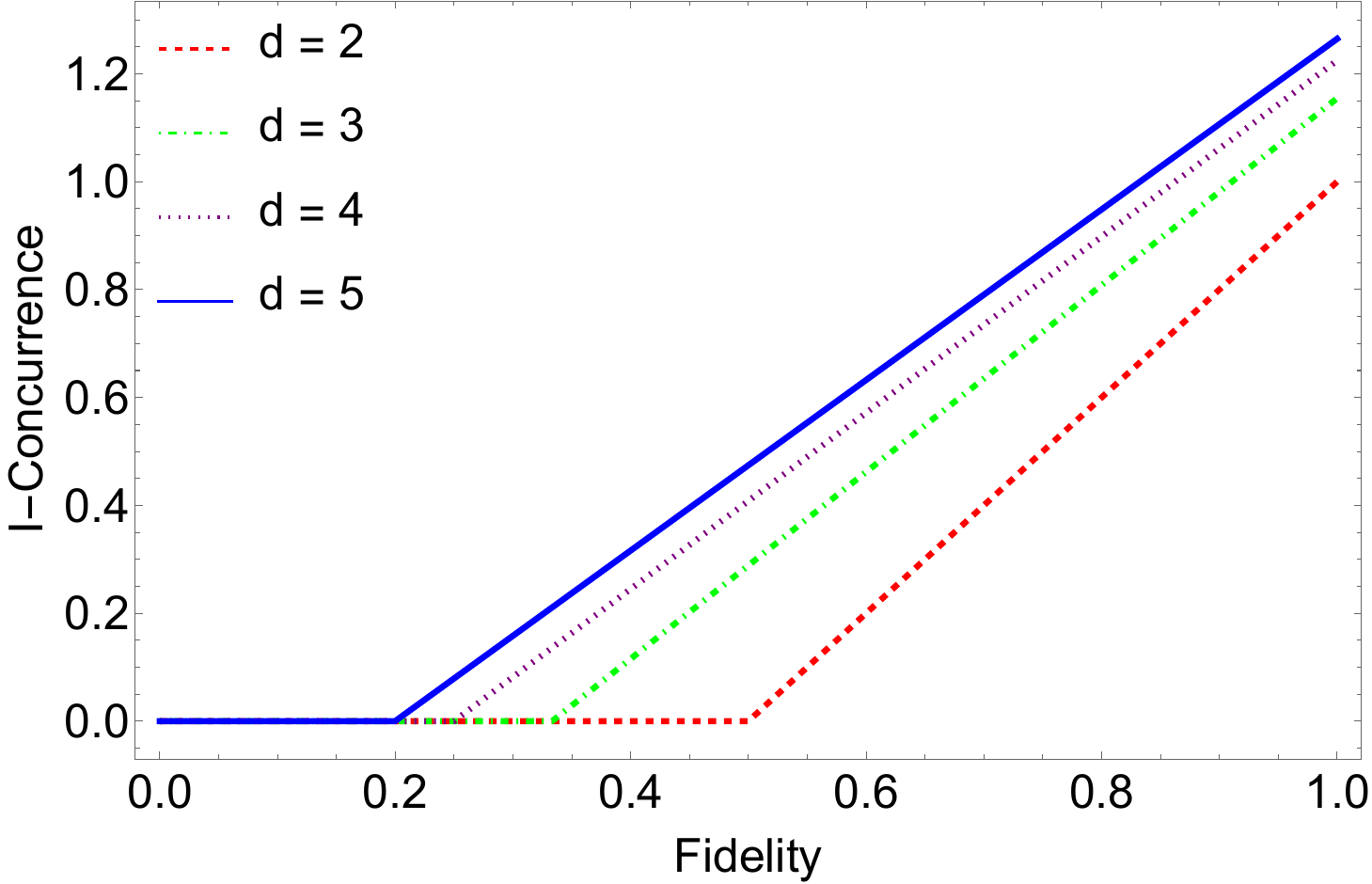}
        \hfill
        \caption{(Color Online) Variation of I-concurrence with fidelity for quantum systems of dimensions $d=2,3,4,5$. The curves highlight the dimensional dependence of the entanglement-fidelity relationship, illustrating nonzero I-concurrence with the least fidelity trend as the dimension increases.}
        \label{fig:Iconc-fidelity}
    \end{figure}

     Figure \ref{fig:Iconc-fidelity} illustrates the relationship between fidelity and I-concurrence for final quantum systems of varying dimensions ($d=2,3,4,5$). Here, increasing fidelity means the noise of the system gradually decreasing and system approaching to noise free behavior.  Thus, higher fidelity corresponds to lower noise and stronger entanglement preservation during the swapping process.
    The fidelity is plotted along the $x$-axis, while I-concurrence, an indicator of quantum entanglement, is represented on the $y$-axis. The curves show how entanglement changes as fidelity varies, with each curve corresponding to a specific system dimension.

    At lower dimensions (e.g., $d=2,3$), the relationship appears smoother and more inclined, whereas higher dimensions ($d=4,5$) exhibit increasingly linear behavior. The blue line shows that we get nonzero I-concurrence even for low fidelity values (around 0.2) in higher-dimensional systems; however, in lower-dimensional systems, the I-concurrence becomes nonzero only at higher fidelities. This observation reflects how dimensionality affects the contribution of $\beta^{\prime}$ to $F$, making high-dimensional systems more tolerant to noise and capable of retaining quantum correlations even when the fidelity is relatively low. The figure highlights the significant impact of dimensionality on the interplay between these two key quantum properties, offering insights into the rich structure of entanglement in high-dimensional spaces.

    The negativity is a measure of quantum entanglement that quantifies the degree of violation of the positive partial transpose (PPT) criterion for separability. To compute the entanglement in terms of negativity, we now consider the partial transpose on subsystem A. After partial transpose of $\rho_{AD}^{N}$, we get
    \begin{equation}\label{pt-isotropic}
    \left(\rho_{AD}\right)^\Gamma=\beta^{\prime}\left(\left|\Phi\right\rangle
     \left\langle{\Phi} \right|\right)^{\Gamma}+(1-\beta^{\prime}) \frac{\mathbb{I}_{d^2}}{d^2},
    \end{equation}
    where $\Gamma$ represents partial transpose with respect to qudit A and $\left|\Phi\right\rangle=\left(\sum_{k=0}^{d-1} e^{-2 \pi i \mu k / d}\left|k\right\rangle\left|k\ominus\nu\right\rangle\right)$. The term $\left(\left|\Phi\right\rangle
     \left\langle{\Phi} \right|\right)^{\Gamma}$ has the eigenvalues $\frac{1}{d}$ and $-\frac{1}{d}$. The first eigenvalue repeats $d(d-1)$ times and the second eigenvalue repeats $d$ times. It means the $\left(\rho_{AD}\right)^\Gamma$ has eigenvalues $\frac{1-\beta^{\prime}}{d^2}+\frac{\beta^{\prime}}{d}$ (repeats $d(d-1)$ times) and $\frac{1-\beta^{\prime}}{d^2}-\frac{\beta^{\prime}}{d}$ (repeats $d$ times). If $\beta^{\prime}>\frac{1}{d+1}$, then the eigenvalue $\frac{1-\beta^{\prime}}{d^2}-\frac{\beta^{\prime}}{d}$ becomes negative. As negativity is the absolute sum of the negative eigenvalues of $\left(\rho_{AD}\right)^\Gamma$, so
    \begin{equation}\label{Nty-isotropic}
    \mathcal{N}\left(\rho_{AD}\right)= \max\left(0,\frac{\beta^{\prime}\left(d+1\right)-1}{d}\right).
    \end{equation}
    We know that fidelity is given by $F=\beta^{\prime}+\frac{1-\beta^{\prime}}{d^2}$. So, the negativity \cite{lee2003convex} in terms of fidelity is
    \begin{equation}\label{iso-fneg}
    \mathcal{N}\left(\rho_{AD}\right)= \begin{cases}0, & F \leqslant \frac{1}{d}, \\ \frac{F d -1}{d-1}, & \frac{1}{d} < F \leqslant 1.\end{cases}
    \end{equation}
    \begin{figure}[t]
         \centering
        \includegraphics[width=0.48\textwidth]{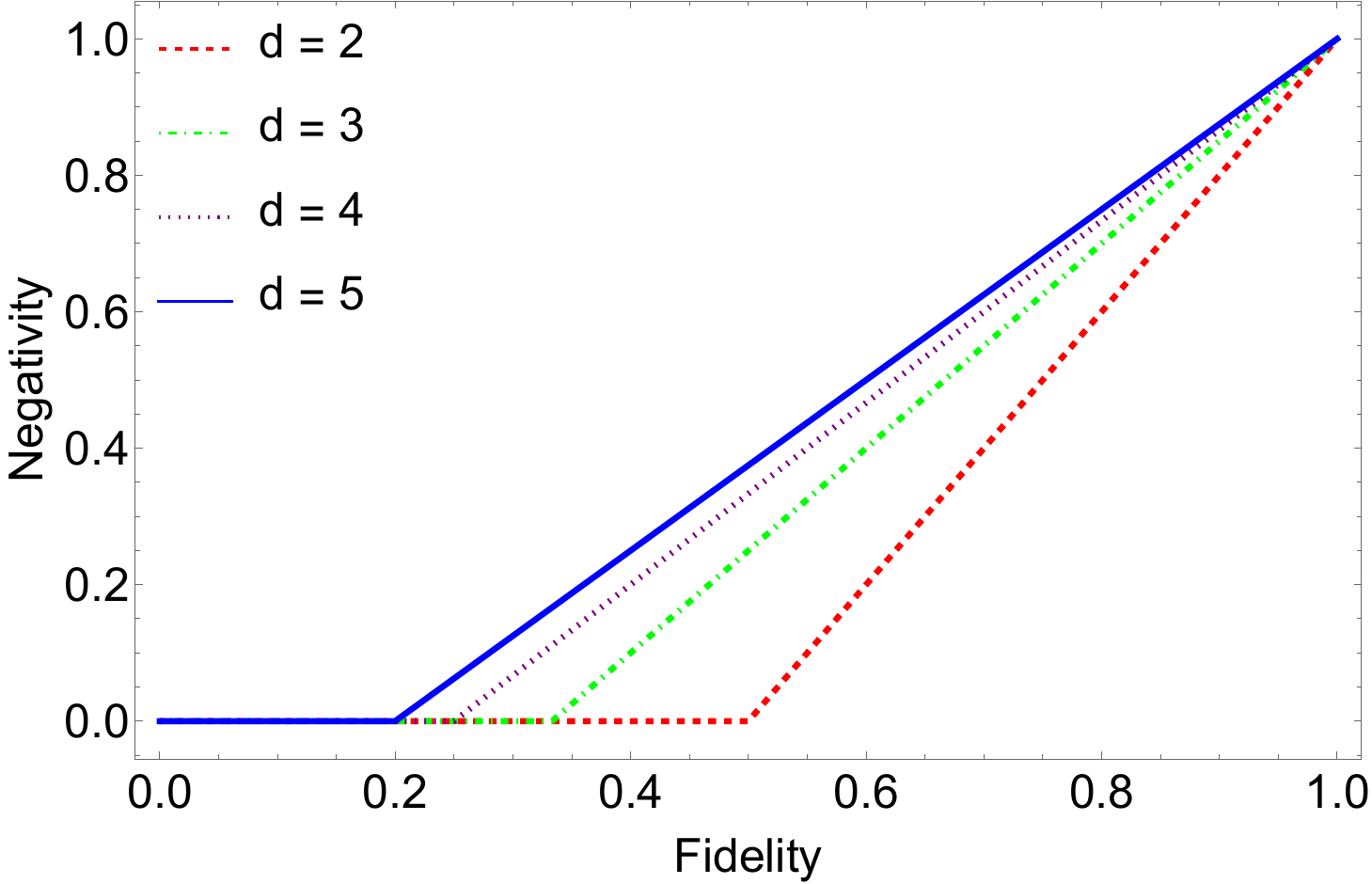}
        \hfill
        \caption{(Color Online) Relationship between fidelity and negativity for isotropic quantum states across different dimensions ($d=2, 3, 4, 5)$. The curves describe nonzero negativity with the least fidelity trend as the dimension increases. All the curves here converge to $\mathcal{N}\left(\rho_{AD}\right)=1$ for $F=1$, independent of dimensions.}
        \label{neg-fidelity}
    \end{figure}
     Figure \ref{neg-fidelity} effectively shows how the relationship between fidelity and negativity depends on the system's dimension. The negativity curve is steep for low dimensions $(d=2,3)$, indicating that a small increase in fidelity results in a noticeable rise in entanglement.
    But for dimensions $(d=4,5)$, the negativity curve flattens. This indicates that achieving high entanglement requires a higher fidelity.
    It means that as the dimension increases, the relationship between fidelity and negativity becomes progressively less steep, indicating reduced sensitivity of negativity to fidelity at higher dimensions. Negativity increases monotonically with fidelity for each dimension, but approaches a smaller nonzero value of fidelity as dimension $d$ increases.
    For each dimension, the negativity starts at zero for $F\leq 1/d$ and begins to rise only when $F>1/d$. This reflects the separability criterion for isotropic states, where entanglement is only present if $F>1/d$.

    Figures \ref{fig:Iconc-fidelity} and \ref{neg-fidelity} compare the dependence of I-concurrence and negativity on fidelity for different Hilbert-space dimensions. In both measures, entanglement remains zero below a dimension-dependent fidelity threshold and then increases linearly with fidelity once this threshold is exceeded. As the dimension increases from $d=2$ to $d=5$, the threshold shifts toward lower fidelity values. Consequently, for a fixed fidelity above the threshold, larger dimensions exhibit larger values of both I-concurrence and negativity. Although the two measures display the same qualitative behavior and identical ordering with respect to the dimension, they differ in their normalization. As shown in Fig. \ref{neg-fidelity}, negativity is normalized to the interval $[0, 1]$ for all dimensions, providing a dimension-independent scale for comparing entanglement. In contrast, the maximum value of I-concurrence increases with the system dimension, as illustrated in Fig. \ref{fig:Iconc-fidelity}, where it exceeds unity for $d>2$. Therefore, I-concurrence reflects the intrinsic dimension dependence of the entanglement measure, whereas negativity provides a bounded and normalized quantifier that facilitates direct comparison across different dimensions.

\section{Conclusion}\label{conclusion}
 In this work, we have investigated entanglement swapping in arbitrary-dimensional quantum systems and analyzed the resulting entanglement distribution using I-concurrence and negativity as quantitative measures. Our results show that the amount of swapped entanglement generally increases with the dimension of the constituent systems, highlighting the potential of high-dimensional entangled states for quantum communication tasks. We have also compared the behavior of I-concurrence and negativity in the swapped states and found that I-concurrence provides a more pronounced characterization of the dimensional dependence of the generated entanglement. Indeed, the dimensional scaling exhibited by I-concurrence and negativity differs significantly, with I-concurrence showing a stronger dependence on the Hilbert-space dimension of the swapped state.

In addition, we examined the influence of the initial entangled states and measurement outcomes on the entanglement-swapping process. To assess the protocol under realistic conditions, we extended our analysis to noisy systems described by isotropic states and studied the dependence of the swapped entanglement on both fidelity and system dimension. The results indicate that higher-dimensional systems can retain nonzero entanglement over a broader range of noise parameters than lower-dimensional systems.

Therefore, our work provides a quantitative characterization of entanglement swapping in high-dimensional quantum systems and clarifies how the dimensionality of the Hilbert space influences both the generated entanglement and its robustness against noise. These findings contribute to the understanding of high-dimensional entanglement distribution and may be relevant for future quantum communication and networking protocols.

\vspace{0.3cm}

\noindent \textbf{Author contributions:}
S.M.Z., C.S., and K.N. contributed to the conceptualization of the study, performed the calculations, interpreted the results, and prepared the original draft of the manuscript. S.H. contributed to the methodology and investigation and participated in reviewing and editing the manuscript.
All authors thoroughly checked the manuscript, discussed the results, and approved the final version of the manuscript.\\
\\
\textbf{Data~availability:}
All data that support the findings of this study are included within the article.\\
\\
\textbf{Competing~interests:}
The authors declare that they have no competing interests.\\
\\
\textbf{Research~funding:}
 This research did not receive funding.

    \bibliography{esse.bib}
\end{document}